\documentclass[amsmath,amssymb,aps,prb,twocolumn]{revtex4-2}
\usepackage[utf8]{inputenc}
\usepackage{xcolor}
\usepackage{eqnarray,amsmath}
\usepackage[title]{appendix}
\usepackage{comment}
\usepackage{graphicx}
\usepackage{dcolumn}
\usepackage{bm}
\usepackage{braket}
\usepackage{comment}

\begin{document}

\title{Geometrical Confinement of Magnetic Domains and Domain Walls on Two-Dimensional Cylindrical-Shaped Ferromagnetic Surfaces}

\author{Gabriel Martínez-Carracedo$^{1,2}$}
\author{Aurelio Hierro-Rodríguez$^{1,2}$}
\author{Amador García-Fuente$^{1,2}$}
\author{Jaime Ferrer$^{1,2}$}

\affiliation{
$^1$Departamento de Física,  Universidad de Oviedo,  33007 Oviedo, Spain\\
$^2$Centro de Investigación en Nanomateriales y Nanotecnología, Universidad de Oviedo-CSIC, 33940, El Entrego, Spain\\
}

\begin{abstract}
The rolling of two-dimensional (2D) van der Waals (vdW) magnetic monolayers into cylindrical architectures provides an ideal platform for confining magnetic domain walls (DWs) without edge-pinning effects. Here, we present a combined analytical and numerical investigation of DW confinement and interaction in ferromagnetic (FM) nanotubes. Using an analytical Heisenberg model on a 2D rectangular lattice, we show that transversal curvature introduces an azimuthal exchange energy penalty that intrinsically compresses the DW width. This model demonstrates excellent quantitative agreement with atomistic Landau-Lifshitz-Gilbert (LLG) spin dynamics simulations. By extending our study to realistic CrSBr (CSB) nanotubes, we model a bound state of two interacting DWs arising from a balance between long-range dipolar attraction and short-range exchange repulsion. Micromagnetic simulations reveal a non-monotonic evolution of the inter-wall distance with the tube perimeter.  These findings highlight vdW magnetic nanotubes as robust platforms for stable domain confinement in quasi-one-dimensional spintronic architectures.

\end{abstract}


\maketitle


\section{Introduction}
2D vdW magnetic materials have emerged at the forefront of condensed matter physics owing to their remarkable and highly tunable properties \cite{Huang2017,Gong2017,Wang2022,Li2022,2pff-xy6n}. Compounds such as CSB, CrI$_3$, CrGeTe$_3$, Fe$_3$GeTe$_2$ and NiPS$_3$ have attracted significant interest, as their magnetic properties can be effectively manipulated via applied strain, interlayer sliding, twisting, or magnetoelectric fields \cite{Fei2018,https://doi.org/10.1002/adma.202415774,10.1103/PhysRevLett.133.246703,Sivadas2018,Hu2020,PhysRevB.110.184406,Cheon2025}. The fundamental drive to understand these systems is largely motivated by their potential integration into next-generation spintronic and memory devices \cite{RevModPhys.96.015005,Borras2024,Zhang2024}, where the external control of their magnetic properties allows for extended functionalities beyond conventional spintronics. Among such potential applications, magnetic racetrack memories represent a highly promising architecture, encoding data within magnetic DWs that are driven by spin currents along magnetic tracks \cite{Gu2022,Tomasello2014,PhysRevApplied.17.064019}.\\ 

The inherent mechanical flexibility of these vdW monolayers suggests the possibility of their reconfiguration into cylindrical architectures (nanotubes) which may be conceptually envisioned as a 2D sheet rolled into a seamless, hollow geometry \cite{caha2026,Lan2025,Rao2013,Li2025}. Within the continuum framework, micromagnetic models have successfully described magnetization reversal modes and DW dynamics in magnetic nanotubes \cite{Landeros2007, Landeros2010}, forming part of the rapidly expanding field of curved magnetism \cite{Streubel2016, PhysRevB.111.184419, Dugato2025}. However, despite extensive research on 2D vdW magnets, the physics of magnetic nanotubes derived from these materials remains largely unexplored \cite{curvedcri3,Zhang2025,Huang2026,ulAhmad2026}.
In this context, vdW magnetic nanotubes could serve as promising candidates for quasi-one-dimensional racetrack memory devices \cite{Guan2025}. Their inherently cylindrical geometry is expected to facilitate the robust confinement and seamless propagation of magnetic domains along the tube axis, offering a potential route to circumventing the deleterious edge-pinning effects that typically plague conventional flat nanostripes.\\ 

Beyond these geometric and architectural advantages, the transition from a planar flat sheet to a cylindrical geometry fundamentally alters the topology and the underlying symmetry groups of the magnetic system. The imposition of periodic boundary conditions (PBC) along the azimuthal direction inherently quantizes the allowed magnetic textures, endowing the DWs with a topological protection that is strictly tied to a non-zero winding number \cite{nakahara}. This topological constraint severely restricts the continuous unwinding of the magnetic texture into a uniform FM state, thereby providing remarkable robustness against thermal fluctuations and structural defects. Furthermore, the geometric rolling process explicitly breaks the continuous two-dimensional translational symmetry, reducing the system to a symmetry group governed by one-dimensional (1D) axial translations and discrete $C_n$ azimuthal rotations\\

It is worth noting that DWs within these FM nanotube surfaces constitute inherently metastable configurations. Consequently, a primary objective of this work is to systematically investigate the precise conditions under which this metastability ensures sufficient stability to effectively confine magnetic domains.\\

In this work, we focus on the study of magnetic domains and DWs confined within cylindrical FM surfaces. To this end, Section \ref{rectlatt} presents an analytical investigation of DWs in a two-dimensional rectangular lattice rolled into a nanotube geometry. Subsequently, Section \ref{CSB_tube} incorporates the material-specific exchange, anisotropy, and long-range dipolar parameters of CrSBr to simulate multi-domain states, demonstrating the robust spatial confinement of magnetic domains whose equilibrium length along the $z$-axis is directly governed by the nanotube diameter.
\section{Computational method}
To investigate the magnetic configurations of the cylindrical nanostructures, we performed atomistic spin dynamics simulations using the {\sc vampire} software package \cite{Evans_2014}. The time evolution of the atomic spins and the energy minimization were computed by numerically integrating the LLG equation. For all calculations, each site was assigned a Cr atomic spin moment of $\mu_S = 3.0\ \mu_B$. Furthermore, a Gilbert damping parameter of $\alpha = 0.1$ was employed and dipole-dipole interactions were included.

The simulation protocols were specifically tailored for the two distinct geometries studied in this work. For the fundamental analysis of the rectangular lattice mapped onto a cylinder (Sec. \ref{rectlatt}), the system was constructed using Open Boundary Conditions (OBC) along the longitudinal $z$-axis. To nucleate a single DW , the initial magnetic state was straightforwardly divided into two distinct halves: the lower section was magnetized antiparallel to the $-z$-direction (spin-down), while the upper section was set parallel (spin-up). 

Conversely, to accurately capture the interactions between two confined DWs in the realistic CSB nanotubes (Sec. \ref{CSB_tube}), a more sophisticated methodology was required. PBCs were rigorously applied along the $z$-direction to emulate a quasi-one-dimensional continuous system and eliminate spurious edge-pinning effects. The initial state was engineered by defining longitudinally alternating domains (spin-up, spin-down, spin-up) separated by two finite-width transition regions. Within these interfaces, the local magnetization was azimuthally discretized into uniform sectors to artificially enforce specific chiralities (left- or right-handed) for each approaching DW. Crucially, in addition to the short-range exchange and uniaxial magnetocrystalline anisotropy, macroscopic dipole-dipole (magnetostatic) interactions were explicitly included in these CSB simulations. The evaluation of this dipolar field is essential to physically capture the long-range attractive forces governing the two-wall bound state.

For both theoretical setups, the systems were allowed to dynamically relax over an extended simulation timeframe of up to 100 ns. This prolonged relaxation protocol ensures that the delicate energetic balance-whether governed strictly by curvature-induced exchange penalties or by the competition between dipolar attraction and exchange repulsion-converges into a robust metastable bound state. While the global energy minimum of the system is fundamentally a uniform FM single domain, this extensive relaxation guarantees that the system successfully bypasses transient dynamic artifacts and settles into the local minima that stabilize the confined DWs.
\section{Results}

\subsection{Rectangular Lattice on a Cylindrical FM Surface}
\label{rectlatt}
\begin{figure}[h]
    \centering
    \includegraphics[width=1\columnwidth]{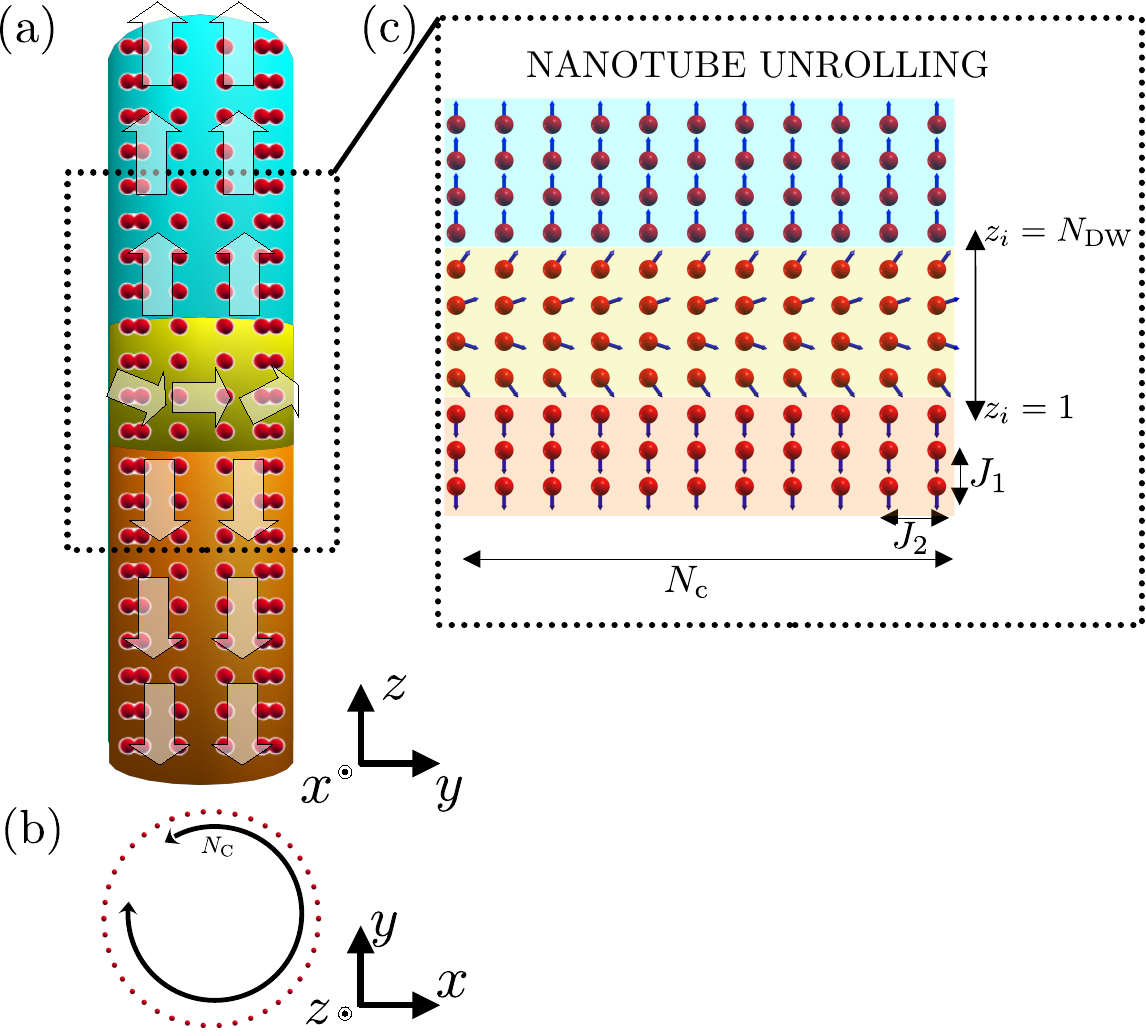}
    \caption{Schematic representation of a Néel DW separating magnetic domains on a rectangular lattice wrapped into a cylindrical geometry. (a) Longitudinal view of the cylinder, where the yellow region highlights the azimuthal rotation of magnetization inside the DW. Cyan (orange) regions represents the spin up (down) domains. (b) Cross-sectional view of the cylinder, where $N_c$ is the number of atoms forming the circumference. (c) Unrolled 2D lattice mapping of the nanotube section enclosed by the dashed box, where blue arrows represent the local spins, $N_{\text{DW}}$ defines the DW width along the $z$-axis, and $J_1$ ($J_2$) denotes the longitudinal (azimuthal) exchange coupling constant.}
    \label{nanotube_TB}
\end{figure}

In this subsection we establish a fundamental baseline by formulating an analytical Heisenberg model on an idealized 2D rectangular lattice rolled into a cylinder. This allows us to isolate how azimuthal exchange curvature penalties intrinsically compress the DW width, validating our analytical scaling laws against atomistic LLG spin dynamics simulations. We investigate the magnetic behavior of  DWs  within a nanotube geometry constructed from a rectangular lattice as depicted in Fig. \ref{nanotube_TB} (a). The system is described by an Heisenberg Hamiltonian for unit spins localized at site $i$, $\mathbf{e}_i=(e_i^x,e_i^y,e_i^z)$, incorporating FM isotropic exchange interactions up to the second-nearest neighbors ($J_1$ and $J_2$). The isotropic exchange $J_1$ couples first-nearest neighbors along the $z$ direction, while $J_2$ couples nearest neighbors along the azimuthal direction (see Fig. \ref{nanotube_TB}(b)). To define the ground-state orientation, a uniaxial anisotropy $K$ is imposed, constraining the magnetic easy axis along the longitudinal direction of the cylinder. The resulting Hamiltonian is given by
\begin{equation}
    H=1/2\sum_{<ij>}J_{ij}\,\mathbf{e}_i\cdot\,\mathbf{e}_j+\sum_i K (e_i^z)^2
    \label{heisengerg_rectangular}
\end{equation}

We investigate the energetic cost of forming a Néel-type DW with axial symmetry. Consequently, the spin orientations are restricted to the tangent planes of the cylindrical nanostructure, possessing only longitudinal and azimuthal components. The DW is shown in Fig. \ref{nanotube_TB}(a). To this end, we evaluate separately the three contributions to the energy formation of the DW, $\delta E^{J_1}$, $\delta E^{J_2}$, and $\delta E^K$. We chose to parametrize the spin vectors using spherical coordinates $\mathbf{e}_i\left(\theta_i,\phi_i\right)=(\text{sin}\,\theta_i\,\text{cos}\,\phi_i\,,\text{sin}\,\theta_i\,\text{sin}\,\phi_i\,,\text{cos}\,\theta_i\,)$ where $\theta$ is measured respect to the $z$-axis and $\phi$ respect to the $x$-axis.\\
Isotropic exchange $J_1$ couples nearest neighbors along the $z$ axis. Spins coupled by $J_1$ have same $\phi$ angle and it only changes $\theta$ along the $z$ axis according to 
\begin{equation}
\theta_i=\pi-\frac{\pi(z_i-1)}{N_{\text{DW}}-1}\hspace{3mm}\text{ for  }\hspace{3mm}z_i=1\dots N_{\text{DW}},
    \label{thetai}
\end{equation}
where $N_{\text{DW}}$ is the number of spins along the DW.\\The energy of a bond coupled by longitudinal exchange $J_1$ is $E_{\text{bond}}^{J_1}=J_1\text{cos}\,\left(\theta_{i+1}-\theta_{i}\right)\simeq J_1\,\left(1-\delta\theta^2/2\right)$ where $\delta\theta=\theta_{i+1}-\theta_{i}=\pi/(N_{\text{DW}}-1)$. So the energy cost from the FM state is $\delta E_{\text{bond}}^{J_1}=-J_1\delta\theta^2/2$. The total contribution of $J_1$ to the DW energy is simply $\delta E_{\text{bond}}^{J_1}$ times $(N_{\text{DW}}-1)$ times $N_c$ so it reads 
\begin{equation}
\delta E^{J_1}=-\frac{J_1\pi^2N_c}{2(N_{\text{DW}}-1)}.
    \label{deltaEj1}
\end{equation}
Azimuthal isotropic exchange $J_2$ couples neighbors with same height $z$, this means that $\theta$ remains constant for spins coupled by $J_2$. As in the previous case, we focus on a bond of a ring at height $z_i$ so $E_{\text{bond}}^{i,\,J_2}=J_2\,\mathbf{e}_i\left(\theta_i,\phi_i\right)\cdot\mathbf{e}_i\left(\theta_i,\phi_{i+1}\right) =J_2 \left[\sin^2\theta_i\cos\delta\phi+\cos^2\theta_i\right]\simeq J_2 \left[\sin^2\theta_i\left(1-\delta\phi^2/2\right)+\cos^2\theta_i\right] $ where $\delta\phi=\phi_{i+1}-\phi_i=2\pi/N_c$. The energy cost from the FM state is $\delta E_{\text{bond}}^{i,\,J_2}=-J_2\,\delta\phi^2\sin^2\theta_i/2$ and considering all bonds coupled by $J_2$ on the ring at height $z_i$ it is obtained $\delta E_{\text{ring}}^{i,\,J_2}=-J_2 N_c\sin^2\theta_i\delta\phi^2/2$. Summing over all rings within the DW, we obtain $$\delta E^{J_2}=\sum\limits_{i}^{N_\text{DW}}\delta E_{\text{ring}}^{i,\,J_2}=-\frac{1}{2}J_2 \delta\phi^2 N_c\sum\limits_{i}^{N_\text{DW}}\sin^2\theta_i\text{ .}$$  Using the identity $\sum\limits_{i}^{N_{\text{DW}}}\sin^2 \theta_i = (N_{\text{DW}} - 1)/2$ and simplifying, we obtain
\begin{equation}
\delta E^{J_2}=-J_2\pi^2(N_{\text{DW}}-1)/N_c.
    \label{deltaEJ2}
\end{equation}
The contribution of the on-site anisotropy energy for all spins inside the DW is given by $E^{\text{DW}}_K=KN_c\sum\limits_{i}^{N_\text{DW}}\cos^2\theta_i$. Using the identity  
$\sum\limits_{i}^{N_{\text{DW}}}\cos^2 \theta_i = (N_{\text{DW}} + 1)/2$ it reads $E^{\text{DW}}_K=KN_c(N_{\text{DW}} + 1)/2$. Subtracting the energy of the FM state, we obtain
\begin{equation}
\delta E^K=-KN_c(N_{\text{DW}}+1)/2
    \label{deltaK}
\end{equation}
So the energy cost to form the DW described above is given by
\begin{equation}
\delta E_{\text{DW}}=\delta E^{J_1}+\delta E^{J_2}+\delta E^{K}
    \label{deltaEDW}
\end{equation}
By minimizing $\delta E_{\text{DW}}$ with respect to $N_{\text{DW}}$, we obtain the optimal DW length given by
\begin{equation}
N_{\text{DW}}=\sqrt{\dfrac{J_1}{2J_2/N_c^2+K/\pi^2}}+1
    \label{NDW}
\end{equation}

This expression represents a generalization of the standard formula $N_{\text{DW}}\propto\sqrt{J/K}$ for a 1D chain \cite{simon2013oxford} to a 2D curved surface.  Equation (\ref{NDW}) indicates that the DW length increases with $J_1$, analogous to the behavior of a 1D chain, but decreases with $J_2$. The latter effect arises from the cylinder's curvature; the exchange interaction $J_2$ couples spins located in different axial planes, which are non-parallel due to the underlying geometry. Consequently, the curvature induces an exchange energy penalty that tends to compress the DW.\\

\begin{figure}[h]
    \centering
    \includegraphics[width=\columnwidth]{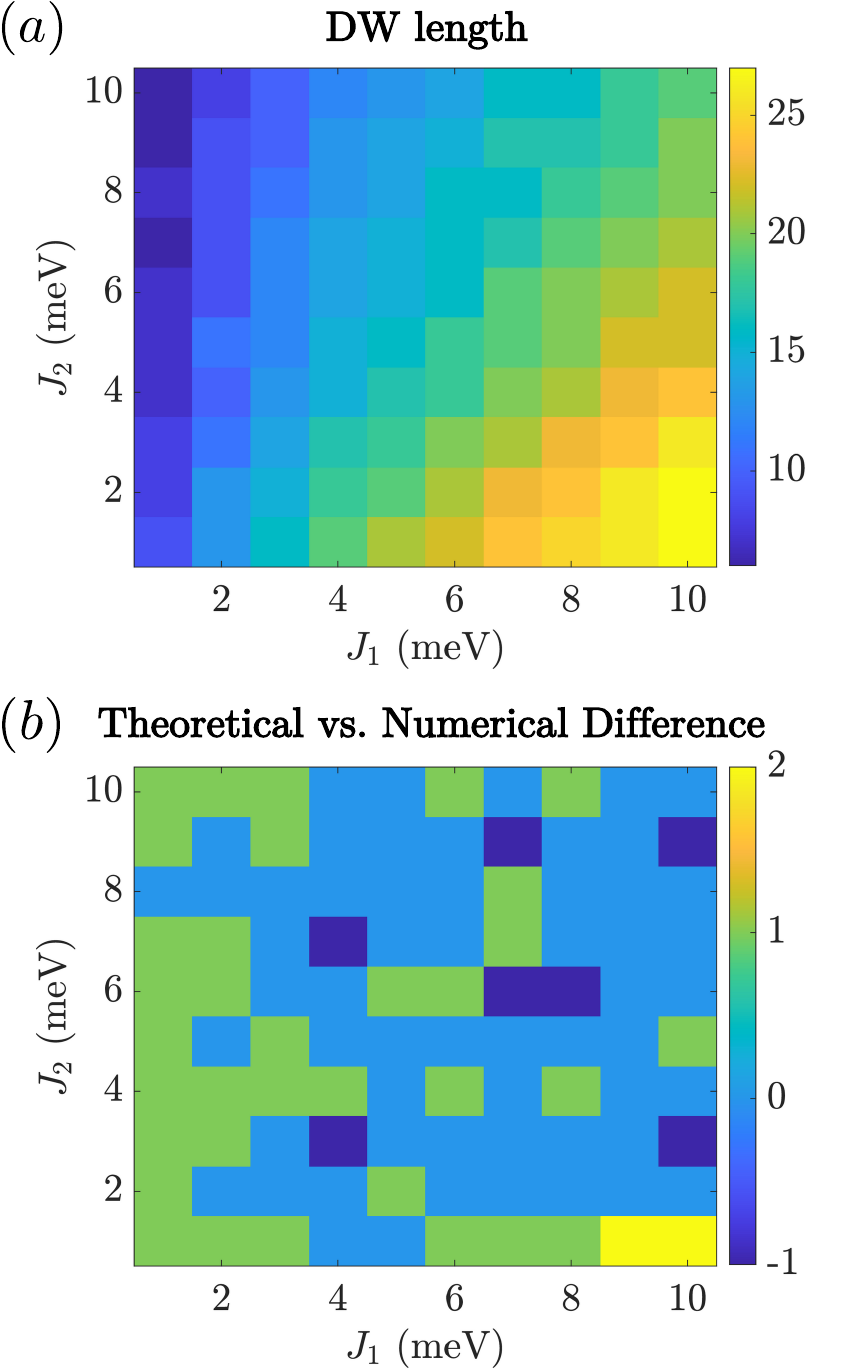}
    \caption{Dependence of the DW length $N_{\text{DW}}$ on the exchange coupling constants. (a) Colormap of the DW length calculated using atomistic simulations ({\sc vampire}) as a function of $J_1$ and $J_2$ for a fixed number of cells $N_c = 30$. (b) Rounded difference between the numerical results from {\sc vampire} and the theoretical predictions from Eq. (\ref{NDW}). The color scale in (b) highlights the regions of maximum deviation between the model and the simulation.}
    \label{square_J1_J2}
\end{figure}

\begin{figure}[h]
    \centering
    \includegraphics[width=\columnwidth]{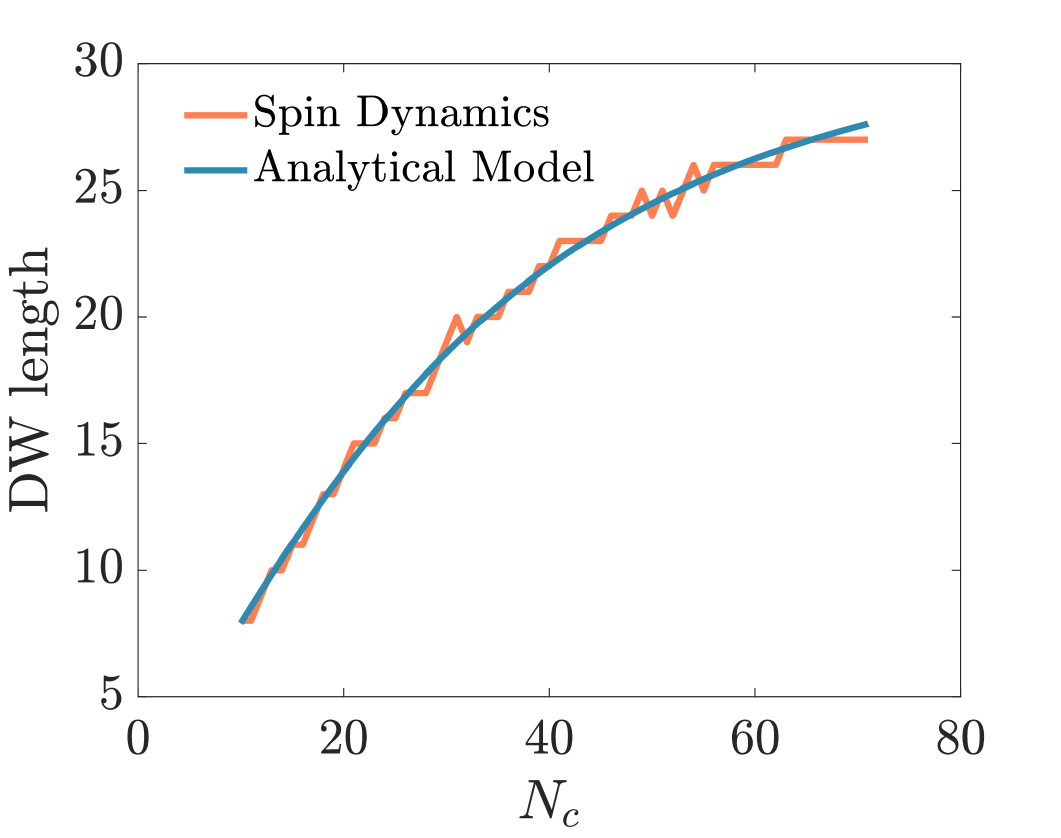}
    \caption{DW length $N_{\text{DW}}$ versus the number of cells $N_c$, with exchange couplings fixed at $J_1 = J_2 = 10$ meV. A comparison between the results obtained from atomistic spin dynamics simulations (orange line) and the analytical model derived in Eq. (\ref{NDW}) (blue line) is shown }
    \label{square_Nc}
\end{figure}

Figs. \ref{square_J1_J2} and \ref{square_Nc} show the agreement between the analytical model given by Eq. (\ref{NDW}) and the {\sc vampire} atomistic simulations for an anisotropy of $K = -0.1$ meV. The DW length was numerically determined from the layer-resolved magnetization profiles (MPs) $\langle e_z(z) \rangle$ extracted from the atomistic spin dynamics simulations, using a threshold criterion of $|\langle e_z \rangle|$ falling below 98.5\% of its saturation value to define the transition region.

The temporal stability of the magnetic domains within the nanotubes can be modeled by the Néel-Arrhenius thermal relaxation model \cite{Neel1949,Brown1963,PhysRevB.104.094433}. For a single domain D1 of length $N_\text{D1}$, the characteristic relaxation time $\tau$ required to thermally overcome the anisotropy energy barrier, $\Delta E_K=K N_c N_\text{D1}$, is given by:
\begin{equation}
\tau = \tau_0 \exp\left( \frac{\Delta E_K}{k_B T} \right),
\label{tau}
\end{equation}
where $k_B T$ is the thermal energy and $\tau_0$ represents the inverse attempt frequency, typically ranging from $10^{-9}$ to $10^{-12}$ s. Eq. (\ref{tau}) allows us to determine the critical domain length for stability
\begin{equation}
    N_\text{D1}=\dfrac{k_BT}{K\,N_{\text{c}}}\text{log}(\tau/\tau_0).
    \label{ND1}
\end{equation}
For fixed parameters $T=300$ K, $K=-0.1$ meV, and $\tau=10^{10}$ s, $N_\text{D1}$ is found to range from 100 to 600 as $N_{\text{c}}$ varies between 20 and 100. Therefore, this guarantees that the domains are robust against spontaneous thermally activated magnetization reversal.
\subsection{Cylindrical CSB Nanostructure}
\label{CSB_tube}
\begin{figure}[h]
    \centering
    \includegraphics[width=\columnwidth]{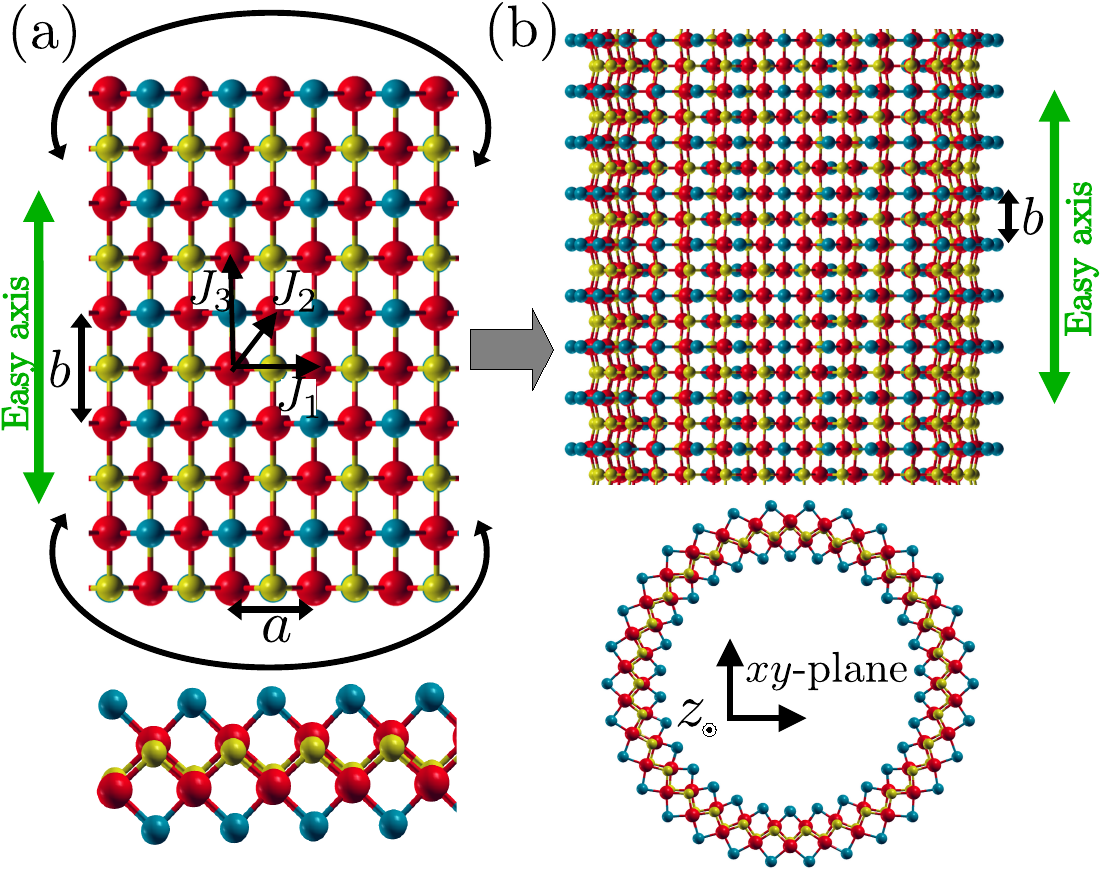}
    \caption{(a) Top and side views of a CSB monolayer. The lattice structure is defined by parameters $a$ and $b$, where $b$ corresponds to the magnetic easy-axis direction. The intralayer exchange couplings $J_1, J_2$, and $J_3$ between first, second, and third nearest neighbors are indicated. (b) Schematic of a CSB nanotube formed by rolling up the monolayer with the tube axis parallel to the $b$ direction. The top and bottom panels display the side view and the cross-section of the cylinder, respectively. }
    \label{CSB_NT}
\end{figure}

In this subsection we extend the framework shown in Sec. \ref{rectlatt} to realistic vdW monolayer CSB nanotubes. By incorporating material-specific exchange parameters, easy-axis magnetocrystalline anisotropy along the tube axis, and long-range magnetostatic dipolar interactions, we investigate the stability of multi-domain configurations and the equilibrium separation between confined DWs in these quasi-one-dimensional systems.
CSB is a vdW material that can be exfoliated down to the 2D limit \cite{https://doi.org/10.1002/adma.202307195}. The monolayer is FM, with the easy axis oriented along the $b$-direction (see Fig. \ref{CSB_NT}) and a Curie temperature around 140 K \cite{Lee2021}. For the atomistic modeling of the CSB nanotubes, we adopt the experimental in-plane lattice parameters of the monolayer, specifically $a = 3.507\ \text{\AA}$ and $b = 4.749\ \text{\AA}$ \cite{Lopez-Paz2022}. In this work, we propose to exploit this specific magnetic anisotropy to construct a magnetic nanotube where the easy axis is parallel to the cylinder axis \cite{Zhang2025}. This configuration enables the confinement of magnetic domains of varying sizes along the length of the tubes.\\

To model the magnetic behavior of these nanotubes, we consider isotropic exchange couplings up to the third nearest neighbors, adopting the values reported for the monolayer \cite{Dey_2025}. Although curvature effects may become relevant at small radii, we assume as an approximation that the variation in these couplings is sufficiently small to avoid significant qualitative changes in our results. The easy-axis anisotropy is set to $K = -0.1$ meV, a value consistent with the range reported in previous theoretical studies of this material \cite{Dey_2025,Rudenko2023,D3NR02518G,https://doi.org/10.1002/adma.202415774}.\\

\begin{figure}[h]
    \centering
    \includegraphics[width=0.8\columnwidth]{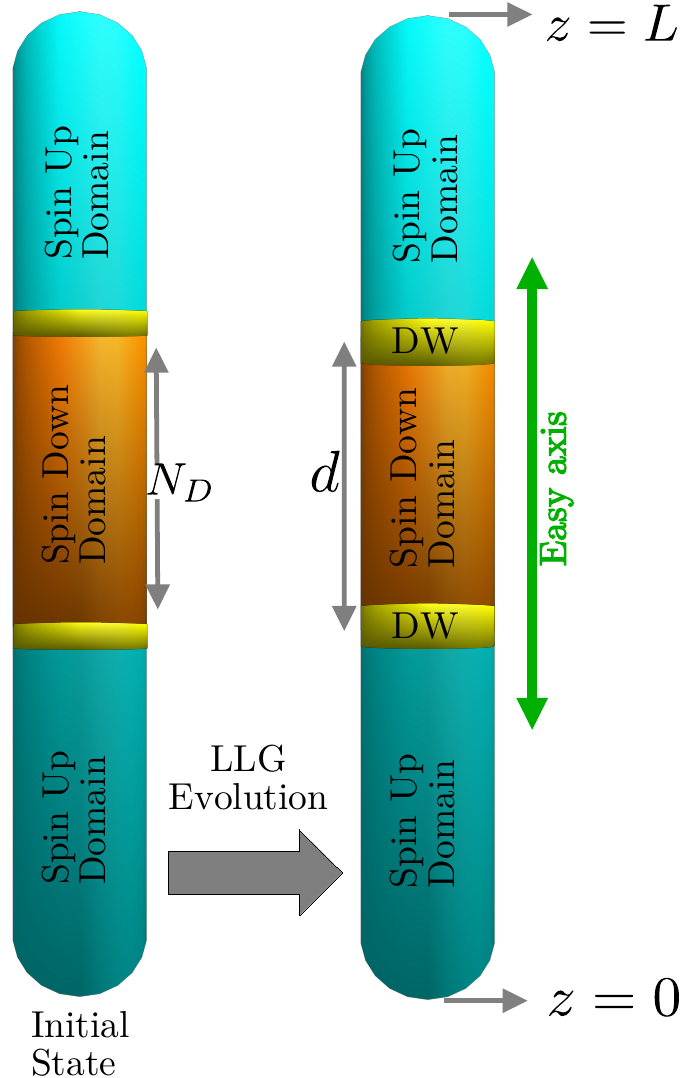}
    \caption{ Schematic illustration of the micromagnetic simulation setup for a $\text{CSB}$ nanotube containing a spin-down domain (orange) sandwiched between two spin-up domains (blue). Starting from an initial state where spins are aligned along the $z$-axis (easy axis), the intermediate regions with $e_z = 0$ are seeded with a specific orientaion to define the desired DW chirality. The system then undergoes LLG evolution until it reaches the equilibrium state, where well-defined DWs (yellow) separated by a distance $d$ are formed.}
    \label{tube_scheme}
\end{figure}

    We investigate the stability of magnetic domains in CSB nanotubes of varying widths and a fixed length of 947.3 Å. Figure \ref{tube_scheme} illustrates the schematic representation of these domains. We initialize the system with a spin configuration characterized by distinct, alternating up-down-up domains. These domains are separated by transition regions where $e_z = 0$, which subsequently evolve to form the DWs. The DWs formed in CSB nanotubes are of the Néel type described in Sec. \ref{rectlatt}, as can be seen in Fig. \ref{nanotube_TB}. This implies that a right- or left-handed chirality is associated with the DW via the right-hand rule. Figure \ref{tube_scheme} shows a CSB nanotube with two DWs separating spin-up and spin-down domains. If both DWs have opposite chirality, there is no topological protection against DW propagation, as a continuous transformation of the magnetization yields a fully FM configuration, causing the DWs to annihilate each other. However, if the chiralities of both DWs are the same, the domains are topologically protected.\\

The interaction between two DWs confined within a quasi-one-dimensional magnetic nanotube can be phenomenologically described by the competition between long-range magnetostatic forces and short-range exchange interactions \cite{kunz2006field}. The mutual interaction may be accurately modeled by considering them as interacting dipoles, leading to an attractive potential energy profile $U_d(d) = -C/d^3$, where $d$ denotes the separation distance between the two DWs. This dipole-dipole potential yields an attractive force whose magnitude is dictated by the parameter $C$, which reflects the internal magnetization profile within the DW structures.\
Conversely, as the DWs approach each other, the spatial overlap of their magnetization tails induces a strong repulsive potential, which we define as $U_e(d) = \sigma_0 e^{-d/\Delta}$. In this expression, $\Delta=N_{DW}b$ represents the characteristic DW width  and $\sigma_0$ is the exchange energy prefactor.\
This fundamental competition between forces gives rise to a stable DW bound state \cite{Thomas2012}. Its characteristic equilibrium distance $d$ is determined by the force balance condition $-\frac{\partial}{\partial d}(U_d + U_e) = 0$, which yields the transcendental equation $d^4 e^{-d/\Delta} = \frac{3C\Delta}{\sigma_0}$. By introducing the dimensionless inter-wall distance $x = d/\Delta$, the stability condition can be elegantly rewritten as $x^4 e^{-x} = \frac{3C}{\sigma_0 \Delta^3} \equiv \lambda$. Here, $\lambda$ emerges as the fundamental dimensionless competition parameter dictating the existence of a stable two-wall system. The explicit dependence of $\lambda$ on the discrete transversal geometry of the nanotube, parametrized by the number of azimuthal cells $N_c$, reveals a complex, non-monotonic evolution of the DW separation. The macroscopic dipolar constant scales with the square of the cross-sectional area, yielding $C \propto N_c^2$, while the exchange penalty along the perimeter dictates $\sigma_0 \propto N_c$. Crucially, the wall width $\Delta$ is not a constant, but is strongly coupled to the transversal curvature as we studied for the square lattice in Fig. \ref{square_Nc}. According to Eq. \ref{NDW} for moderate values of $N_c$, it is expected that the relief of the curvature-induced exchange energy causes $\Delta(N_c)$ to expand rapidly. This substantial broadening of the DWs significantly stretches the spatial range of the exchange repulsion, temporarily overpowering the enhanced dipolar attraction and leading to an initial increase in the physical separation $d$, as observed in micromagnetic simulations. However, in the asymptotic limit of macroscopically wide tubes ($N_c \to \infty$), the DW width saturates to a constant monolayer limit $\Delta_\infty$, dictated solely by the magnetic anisotropy and exchange with no curvature effects. In this macroscopic regime, the dimensionless parameter $\lambda(N_c) \propto N_c / \Delta_\infty^3$ grows linearly without bound. Once $\lambda$ systematically exceeds the critical mathematical threshold for stable roots ($\lambda_c = 256/e^4 \approx 4.69$), the purely quadratic scaling of the dipolar attraction inevitably crushes the exchange repulsion. Upon reaching this asymptotic collapse threshold, the ultimate fate of the system is strictly governed by topology. If the approaching DWs share the same chirality, the conservation of the total winding number forbids any continuous deformation into a collinear state. 

\begin{figure}[h]
    \centering
    \includegraphics[width=\columnwidth]{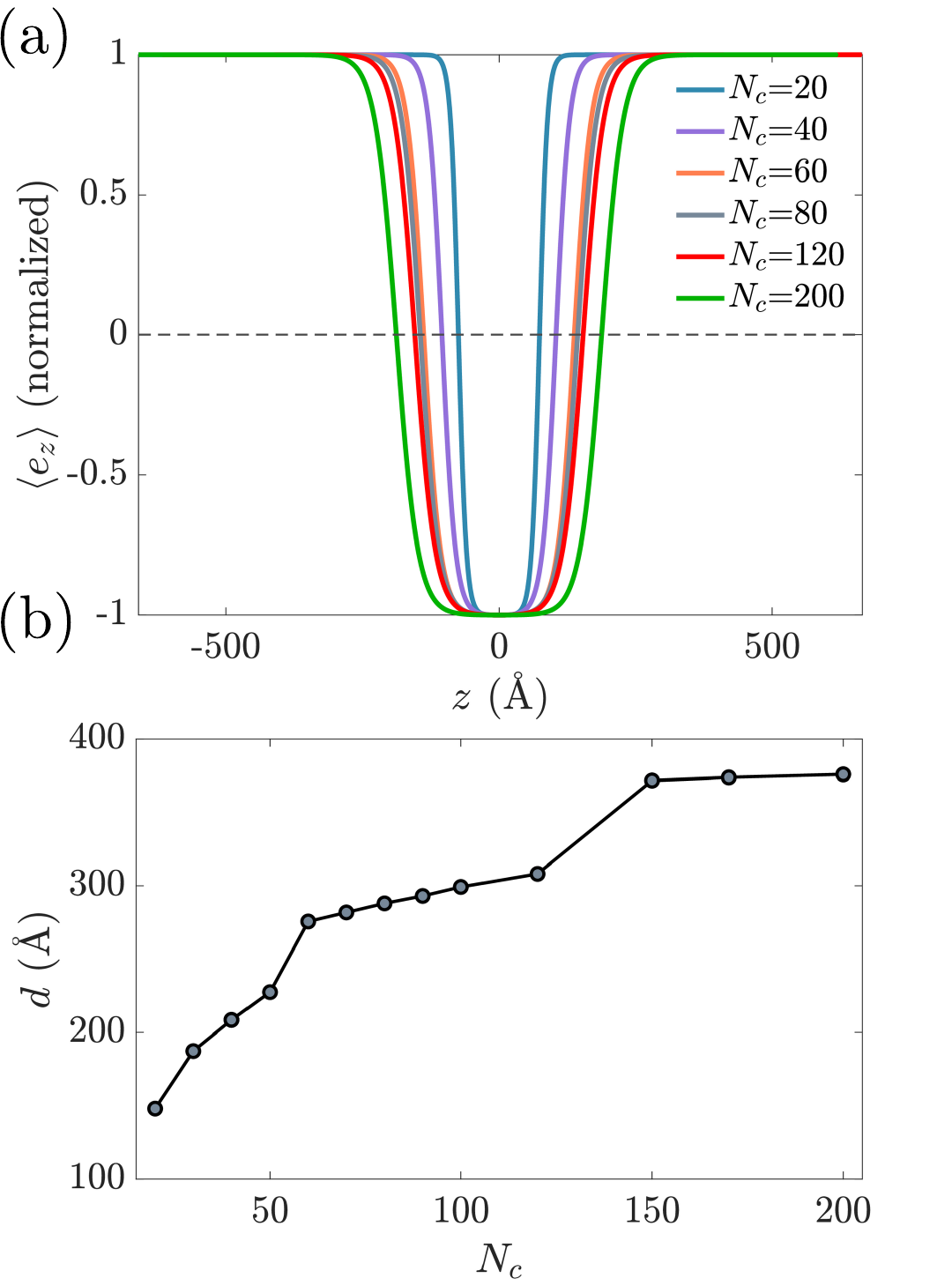}
    \caption{ (a) MPs of a CSB nanotube of length $L = 1275\text{ \AA}$ obtained from LLG dynamics versus the nanotube perimeter $N_c$, for the fixed initial condition $N_D = L/5$, at $T = 0\text{ K}$. (b) Distance $d$ between DWs as a function of $N_c$.}
    \label{magnetization_profile}
\end{figure}

To corroborate the phenomenological model, Fig. \ref{magnetization_profile}  presents the micromagnetic simulation results for the two-wall system as a function of the nanotube perimeter. Figure 6(a) displays the normalized axial MPs ($\langle e_z \rangle$) for increasing nanotube perimeters, ranging from $N_c = 20$ to $200$. The corresponding physical separation $d$ between the DWs is extracted and plotted in Fig. 6(b). In agreement with the theoretical framework, the system exhibits a pronounced, monotonic expansion of the inter-wall distance for moderate values of $N_c$ ($20 \le N_c \lesssim 150$). Within this regime, the relief of the curvature-induced exchange penalty causes the characteristic DW width $\Delta(N_c)$ to expand rapidly. As predicted, this substantial structural broadening stretches the spatial range of the exchange repulsion $U_e(d)$, effectively overpowering the quadratically scaling dipolar attraction $U_d(d)$ and driving the walls further apart. However, as the system transitions toward the macroscopic limit ($N_c \gtrsim 150$), the growth of $d$ noticeably decelerates, leading to a saturation plateau. This behavior precisely reflects the onset of the asymptotic regime where the DW width approaches its constant monolayer limit, $\Delta_\infty$, governed solely by magnetic anisotropy and exchange. In this plateau region, the exchange repulsion no longer gains spatial reach, allowing the dimensionless parameter $\lambda(N_c)$ to grow linearly. The stabilization of $d$ around $N_c = 200$ indicates that the system has entered the macroscopic regime but has not yet systematically exceeded the critical mathematical threshold ($\lambda_c \approx 4.69$); therefore, the purely quadratic scaling of the dipolar attraction has not yet surpassed the short-range exchange repulsion required to drive a structural collapse of the bound state.\\

From a technological standpoint, the physical length of the confined magnetic domains (which effectively encode the data bits in potential racetrack memory applications) is highly tunable via the nanotube geometry. As evidenced by Fig. 6(b), the equilibrium distance of the spin-down domain sandwiched between two spin-up regions can be systematically modulated from 148 to 375 \AA{} for $N_c$ values corresponding to nanotube diameters ranging between $22$ and $220$\AA, respectively. This direct correlation demonstrates that the transverse geometrical confinement provides a versatile and precise mechanism to control the bit size and, consequently, the data density in these quasi-one-dimensional magnetic architectures. It is instructive to put these figures in the context of conventional racetrack memory concepts. In state-of-the-art perpendicular magnetic anisotropy (PMA)-based racetrack, experimentally demonstrated bit lengths have been pushed down to approximately 200 \AA{} \cite{Fukami2013,Parkin2015}. This suggests that magnetic nanotubes could, in principle, reach comparable or higher data densities than conventional PMA racetracks.\\


\section{Summary and Conclusions} 
We have systematically investigated the geometric confinement and stabilization of magnetic domains and DWs within two-dimensional vdW FM nanotubes. By employing an analytical Heisenberg model mapped onto a cylindrical surface, alongside realistic atomistic and micromagnetic simulations parameterized for CSB, we demonstrated that the transition from a planar geometry to a closed cylindrical architecture fundamentally alters the magnetic energetic landscape and topology of the system. 

Our analytical model for a rectangular lattice reveals that the characteristic DW width is tightly coupled to the transversal curvature of the cylinder. Specifically, the azimuthal exchange coupling between adjacent axial planes introduces an energy penalty that intrinsically compresses the DW. This curvature-induced compression modifies the standard one-dimensional DW length, yielding a generalized functional dependence on the discrete transversal geometry ($N_c$). Furthermore, applying a Néel-Arrhenius thermal relaxation model, we verified that domains of moderate lengths within these nanostructures present a remarkable robustness against thermally activated spontaneous magnetization reversal at room temperature.\\

We have established that interacting domain walls can form a metastable bound state at a specific equilibrium distance, dictated by the fine balance between dipolar attraction and exchange repulsion. We modeled their dynamics through the competition between a long-range attractive dipole-dipole potential and a short-range repulsive exchange interaction. Micromagnetic simulations of CSB nanotubes corroborate this framework, revealing a complex, non-monotonic evolution of the inter-wall distance as a function of the tube perimeter. For moderate tube perimeters, the relief of the curvature-induced exchange penalty rapidly expands the DW width $\Delta$, stretching the spatial range of the exchange repulsion and temporarily overcoming the dipolar attraction, which physically drives the walls further apart. 

However, as the system transitions toward the asymptotic planar limit ($N_c \to \infty$), the DW width saturates to a constant value governed by intrinsic anisotropy and exchange. In this macroscopic regime, the purely quadratic spatial scaling of the macroscopic dipolar constant ultimately crushes the exchange repulsion. Upon reaching this collapse threshold, topological constraints determine the resulting magnetic state. If the interacting DWs are topologically protected, the conservation of the winding number prohibits continuous annihilation.\\

Although this study employs CSB as a prototypical system, the physical principles driving the spatial confinement of magnetic domains are essentially universal. The analytical mapping of the rolled rectangular lattice demonstrates that domain wall compression arises fundamentally from the geometric curvature, transcending specific material parameters. Similarly, the energy balance between short-range exchange and long-range dipolar interactions, which dictates the domain length, is a general property of cylindrical boundary conditions. Therefore, our theoretical framework can be directly generalized to the broader class of 2D vdW ferromagnets, offering a universal route to design stable, quasi-one-dimensional magnetic architectures for spintronic applications simply by tuning the macroscopic curvature. Altogether, our findings establish that vdW magnetic nanotubes provide an exceptional platform for confining complex magnetic textures. Furthermore, the interplay between competing magnetic interactions yields a novel self-packing mechanism, offering a versatile route to control bit density in nanoscale storage architectures. The inherent absence of edge effects, combined with the strict topological protection granted by the periodic azimuthal boundary conditions, positions these curved quasi-one-dimensional geometries as highly promising candidates for robust, next-generation racetrack memory architectures and nanoscale spintronic devices.

\section*{Acknowledgements}
G. M.-C., A. G.-F. and J. F. have been funded by Ministerio de Ciencia, Innovación y Universidades, Agencia
Estatal de Investigación, Fondo Europeo de Desarrollo Regional via the grant PID2022-137078NB-I00,
and by Agencia SEKUENS (Asturias) under grant UONANO IDE/2024/000678 with the support of FEDER funds. This project is supported by the TRILMAX Horizon Europe consortium (Grant No. 101159646). A. H.-R. acknowledges the support of Spanish MCIN/AEI/10.13 039/50 110 0011 033/ FEDER, UE under Grant PID2022-13 6784NB and of Agencia SEKUENS (Asturias) under grant UONANO IDE/2024/000 678 with the support of FEDER funds.

\end{document}